\documentclass{jkas}
\def\year{2015} 
\def\month{April} 
\def\volume{48} 
\def\issue{2} 
\def\beginpage{139} 
\def\endpage{154} 
\def\received{September 24, 2014} 
\def\accepted{March 17, 2015} 
\date{Received \received; accepted \accepted}
\usepackage{flushend} 
\usepackage{color}

\def\simlt{\lower.5ex\hbox{$\; \buildrel < \over \sim \;$}}
\def\simgt{\lower.5ex\hbox{$\; \buildrel > \over \sim \;$}}
\def\msol{$M_\odot$}

\title{
Dynamical Evolution of Supernova Remnants\\
Breaking Through Molecular Clouds
}
\author[1]{Wankee Cho}
\author[2]{Jongsoo Kim}
\author[3]{Bon-Chul Koo}
\affil[1]{Department of Physics and Astronomy, Seoul National University, Seoul 151-742, Korea; \email{wkcho@astro.snu.ac.kr}}
\affil[2]{Korea Astronomy and Space Science Institute, 36-1 Hwaam, Yusong, Daejeon 305-348, Korea; \email{jskim@kasi.re.kr}}
\affil[3]{Department of Physics and Astronomy, Seoul National University, Seoul 151-742, Korea; \email{koo@astro.snu.ac.kr}}
\def\corrauthor{
W. Cho
}
\def\runningauthor{
Cho et al.
}
\def\runningtitle{
Dynamical Evolution of Supernova Remnants Breaking Through Molecular Clouds
}
\def\keywords{
Hydrodynamics --- methods: numerical --- ISM: supernova remnants and clouds
}
\def\abstracttext{
We carry out three-dimensional hydrodynamic simulations of the supernova remnants (SNRs) produced inside molecular clouds (MCs) near their surface using the HLL code \citep{har83}. We explore the dynamical evolution and the X-ray morphology of SNRs after breaking through the MC surface for ranges of the explosion depths below the surface and the density ratios of the clouds to the intercloud media (ICM).
    We find that if an SNR breaks out through an MC surface in its Sedov stage, the outermost dense shell of the remnant is divided into several layers. The divided layers are subject to the Rayleigh-Taylor instability and fragmented. On the other hand, if an SNR breaks through an MC after the remnant enters the snowplow phase, the radiative shell is not divided to layers. We also compare the predictions of previous analytic solutions for the expansion of SNRs in stratified media with our one-dimensional simulations.
    Moreover, we produce synthetic X-ray surface brightness in order to research the center-bright X-ray morphology shown in thermal composite SNRs. In the late stages, a breakout SNR shows the center-bright X-ray morphology inside an MC in our results. We apply our model to the observational results of the X-ray morphology of the thermal composite SNR 3C 391.
}
\begin{document}
\jkashead 

\section{Introduction\label{sec:intro}}

Core-collapse supernova remnants (SNRs) often interact with large molecular clouds (MCs). This interaction between SNRs and dense MCs is of considerable interest because it provides an opportunity to study the dynamical and chemical processes associated with strong shocks, e.g., how the MCs affect the evolution of SNRs, how the MCs are disrupted by SN shocks, how the shocks change the abundances of MCs, and how molecules are destroyed or reformed. Since the first discovery of the SNR-MC interaction in the SNR IC 443, \citep{cornett77, denoyer79}, 45 SNRs, which are 16\% of the known Galactic SNRs, have been found to show some evidences for the interaction with MCs according to the compilation by \citet{jiang10}.

   The SNRs interacting with MCs are of particular interest in relation to the \emph{thermal} \emph{composite} or \emph{mixed} \emph{morphology} (MM-) SNRs, which are the SNRs that appear shell-type in radio continuum but emit bright thermal X-ray in its center. In \citet{rho98}, it was surmised that about $\sim$ 25\% of the whole SNRs detected from the $Einstein$ observation belong to this category.
   The center-bright morphology is not consistent with a Sedov-Taylor model \citep{sed46,tay50} where most of the swept-up matter is confined to a dense shell at the boundary. In order to explain the MM-SNRs, several hypotheses have been proposed. Probably the most popular one is the so-called evaporation model of \citet{whi91}. In this model, the ambient medium is clumpy, so there are dense clumps inside the SNR which have survived the passage of the SN shock. Subsequently, these clumps evaporate thereby brightening the X-ray emission. Another hypothesis is based on conduction from a hot interior to cold radiative shell, proposed by \citet{cox99} for the SNR W44.
   The other two hypotheses are fossil thermal radiation from a hot interior after the shell of a remnant cools down \citep{sew85} and projection effect of the SNR exploded outside an MC \citep{pet01}.
   It is worth noting that a good fraction of these MM-SNRs are the SNRs interacting with MCs \citep{rho98,jiang10}. And among the 45 probable SNRs interacting with MCs in the catalog of \citet{jiang10}, 15 SNRs are classified as the MM-SNRs. It is therefore interesting to investigate if an SNR breaking out of an MC can appear centrally-brightened in X-rays.

  Dynamical evolution of SNRs interacting with MCs has been studied numerically by several authors \citep{fal82, ten85, yor89, art91, dohm96, vel01, fer08}. According to their results, the interaction may be divided into three categories.
  First, if an SN explosion ($E_{51}$ = 10$^{51}$ erg as thermal energy) occurs deep inside an MC, e.g., 15 pc below the cloud surface \citep{ten85}, its remnant cannot break out of the MC and ends its life inside the cloud. The possible existence of such buried SNRs has been analytically addressed by \citet{wheeler80} and \citet{shu80}, where the SNRs radiate most of the energy in the infrared energy range.
  Second, if an SN explosion occurs close to the surface of an MC, the SN blast wave can break out of the MC. This breakout phenomenon is characterized by the acceleration of the blast wave and the ejection of cloud matter across the original cloud surface. If the breakout occurs when an SNR is in the Sedov phase, the accelerated blast wave produces a large half-spherical remnant in the low-density intercloud medium (ICM), whereas the blast wave propagating into the dense cloud matter makes another sphere which is well described by the Sedov-Taylor solution and the relation in \citet{cio88}. \citet{dohm96} investigated the early evolution of SNRs produced very near the cloud surface. If an SNR breaks out the surface of an MC during its snowplow phase, the breakout process is expected to be rather complicated with the radiative shell disrupted by the Rayleigh-Taylor (R-T) instability. The overall morphology of the SNR is considerably elongated along the direction perpendicular to the cloud surface.
  Finally, if an SN explodes outside the MC, the cloud is not largely disrupted while the SNR is distorted to a half-sphere. For the SN explosion just outside the cloud, \citet{fer08} presented the results of magneto-hydrodynamic simulations which shows how the reflected wave from the cloud surface moves back to the explosion center. Especially, \citet{ten85} carried out two-dimensional (2-D) simulations with cases belonging to the above three categories and studied the dynamical evolution of SNRs and the cloud disruption efficiency.

  In this paper, we explore the dynamical evolution of breakout SNRs (BO-SNRs) with hydrodynamic simulations with the aim of finding the conditions for the SNRs to show center-bright X-ray morphology.
  \citet{ten85} suggest that additional matter could be supplied by the radiative shell broken by the R-T instability
  so that we may expect the BO-SNRs to show the center-bright X-ray morphology.
  Their simulations with low resolution ($\sim$ 100 computational grids in one dimension), however, are limited for investigating the detailed process of breakout such as the disruption of a radiative shell by the R-T instability.
  So, by performing three-dimensional (3-D) simulations with higher resolution, we could describe the complex structures of the BO-SNRs such as the R-T unstable structures.
  We also synthesize X-ray morphology of BO-SNRs and investigate when they appear centrally-brightened. We apply our result to the SNR 3C 391, which is a prototype of the MM-SNRs.

This paper is organized as follows.
  In Section~2, we introduce the methods used in the numerical simulations with cooling and heating processes.
  In Section~3, we present the results from several models with different depths and density ratios.
  In Section~4.1, we compare the results of numerical simulations to the results of one-dimensional (1-D) spherical experiments, and the semi-analytic solutions of shell and shock by \citet{koo90}.
  Also in Section~4.2, we derive X-ray morphology of the simulated SNRs to discuss the origin of thermal emission inside the MM-SNRs. Finally, a simulated SNR model is compared with the prototypical MM-SNR, 3C 391, in the scope of X-ray characteristics.

\section{Numerical Methods}

\subsection{Governing Equations and Numerical Schemes}

  To follow up the evolution of SNR interacting with MC, we solve the following Eulerian hydrodynamic equations:
    \begin{equation}
    \frac{\partial \rho}{\partial t}+ \nabla \cdot (\rho \textbf{v}) = 0,
    \end{equation}
    \begin{equation}
    \frac{\partial }{\partial t} (\rho \textbf{v})+ \nabla \cdot ( \rho \textbf{v} \textbf{v} )+ \nabla P = 0,
    \end{equation}
    \begin{equation}
    \frac{\partial E}{\partial t} + \nabla \cdot [(E+P)\textbf{v}] =  \Gamma - \Lambda.
    \end{equation}
where the total energy $E$ is defined as  $E ={P}/(\gamma-1) + \rho v^2/2 $ and $\mu$ is the mean molecular weight, $\mu = 14 m_{\rm H} /23$ with 10\% helium fraction by number under fully ionized state. Other symbols have their usual meanings.

  The energy equation must be integrated with effective cooling effect, $\Lambda_{\rm eff}= \Gamma - \Lambda$ to follow the SNR evolution in the snowplow phase. The radiative cooling rate, $\Lambda = n_{\rm e} n_{\rm H} L(T) $ with the cooling function, $L(T)$, and the hydrogen and electron densities, $n_{\rm H}$ and $n_{\rm e}$. In the fully ionized state and with 10\% helium fraction by number, $n_{\rm e} = 1.2 n_{\rm H}$. $L(T)$ involves different
cooling processes as a function of the temperature.
  From 10 K to 10$^4$ K the cooling function of \citet{san02}, which is fitted by a piecewise power-law fit \citep{wol95}, is adopted.
  From 10$^{4}$ K to 10$^{8}$ K, the non-equilibrium cooling curve of \citet{sha76} with the solar abundance is adopted.
  For higher temperature ($T$ $>$ 10$^8$ K), we include thermal bremsstrahlung process in $L(T)$.

The heating rate, $\Gamma= n_{\rm{H}} G(T)$, comes from a process such as photoelectric heating by starlight, composed of the hydrogen density and the heating function, $G(T)$. The heating function is given by $G(T)=1.2 n_{\rm{i}} L(T_{\rm {i}})$ with initial density and temperature, $n_{\rm {i}}$ and $T_{\rm {i}}$.
  Then, the effective cooling effect can be written at the medium as:
   \begin{eqnarray}
       \Lambda_{\rm eff} &=& \Lambda-\Gamma=n_{\rm e}n_{\rm H}L(T)-n_{\rm H}G(T) \nonumber \\
                     &=& 1.2 n_{\rm H} (n_{\rm H} L(T) - n_{\rm i} L(T_{\rm i})).
   \end{eqnarray}
Now the effective radiative cooling can be calculated with the hydrogen number density, $n_{\rm H}$ and the cooling function $L(T)$ with given
initial conditions of $n_{\rm i}$ and $T_{\rm i}$.
Through a model, an MC and the ICM are in thermal equilibrium.

  The hydrodynamic equations are solved using the HLL method \citep{har83}, which solves the Riemann problem in an approximate way to obtain intercell fluxes. Since we do not need a full characteristic decomposition of the equations, the HLL Riemann solvers are straightforward to implement and very efficient. The HLL code is tested both for the Sod problem \citep{sod78} and the SNR evolution in the adiabatic state, which are in good agreement with analytic solutions \citep{shu80}.

The coloring method is the scheme to trace a specific component of the multi-fluid using a Lagrangian tracer variable \citep[see][Equation 1]{xu95}. In addition to the usual hydrodynamic equations, we solve the continuity equation for each component. The density is obtained by multiplying the density of the fluid to a Lagrangian tracer variable. In this paper, we trace the MC and the ICM materials, separately (Figure~4), so that we can verify the origin of the complex structure in the evolved stage.

  For the cooling process, we first calculate the cooling time scale, $\Delta t_{\rm cool} =  {E_{\rm int}^{\rm n}} / { \Lambda _{\rm net} }$, where $E^{\rm n}_{\rm int}$ is the thermal energy and the superscript n represents the n-th time step. We then update the thermal energy using:
      \begin{equation}
            E_{\rm int}^{\rm n+1} = E_{\rm int}^{\rm n} \exp({- 0.5{\Delta t_{\rm dyn}}/{\Delta t_{\rm cool}}})
      \end{equation}
where $\Delta t_{\rm dyn}$ is the dynamical time step set by the Courant condition. Because the above steps are solved before and after the hydrodynamic part, there is 0.5 in Equation (5).

  \begin{figure}[t!]
  \centering
  \includegraphics[trim=45mm 15mm 45mm 25mm, clip, width=80mm]{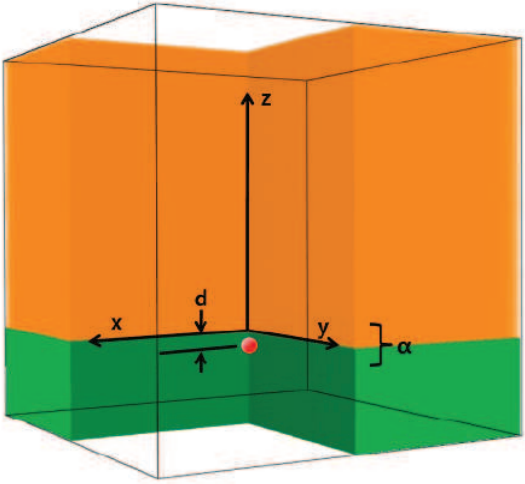}
  \caption{Three-dimensional schematic overview on our models. Three different regions are drawn in different colors: green for a molecular cloud, orange for the intercloud medium and red for the initial SNR. The $x-$ and $y-$axes on the MC surface are set to be perpendicular to the $z-$axis. $d$ is the depth below the cloud surface where an SN explodes and $\alpha$ is the density ratio of the MC to the ICM.}
  \end{figure}

  \begin{figure}[t!]
  \centering
  \includegraphics[width=83mm]{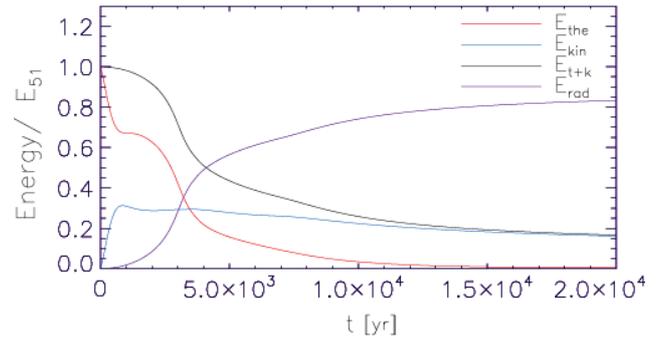}
  \caption{Energy budget of D250 with time. Each solid line represents thermal energy (red), kinetic energy (blue), the total kinetic and thermal energy (black), and energy loss by the radiative cooling (purple), which are normalized by the SN explosion energy of 10$^{51}$ erg. }
  \end{figure}

\begin{table}[t!]
\caption{Model parameters with ejecta of 10 \msol. \label{tbl-1}}
\centering
\begin{tabular}{lrrr}
\toprule
Model & $d$ $^{\rm a}$ & $\alpha$ $^{\rm b}$ & $Res.^{\rm c}$ \\
\midrule
D200  & $2.0$ & $10^3$ & $1/8$ \\
D250  & $2.5$ & $10^3$ & $1/8$ \\
D300  & $3.0$ & $10^3$ & $1/8$ \\
D350  & $3.5$ & $10^3$ & $1/8$ \\
R101  & $2.5$ & $10^1$ & $1/8$ \\
R102  & $2.5$ & $10^2$ & $1/8$ \\
R104  & $2.5$ & $10^4$ & $1/8$ \\
H032  & $2.5$ & $10^3$ & $1/32$ \\
\bottomrule
\end{tabular}
\tabnote{
$^{\rm a}$  The explosion depth $d$ is in the unit of pc.
\\ $^{\rm b}$  The density ratio $\alpha$ is the ratio $n_{\rm MC}/ n_{\rm ICM}$.
\\ $^{\rm c}$  The resolution is in units of pc/grid.
}
\end{table}

  \begin{figure*}[t!]
  \centering
  \includegraphics[width=170mm]{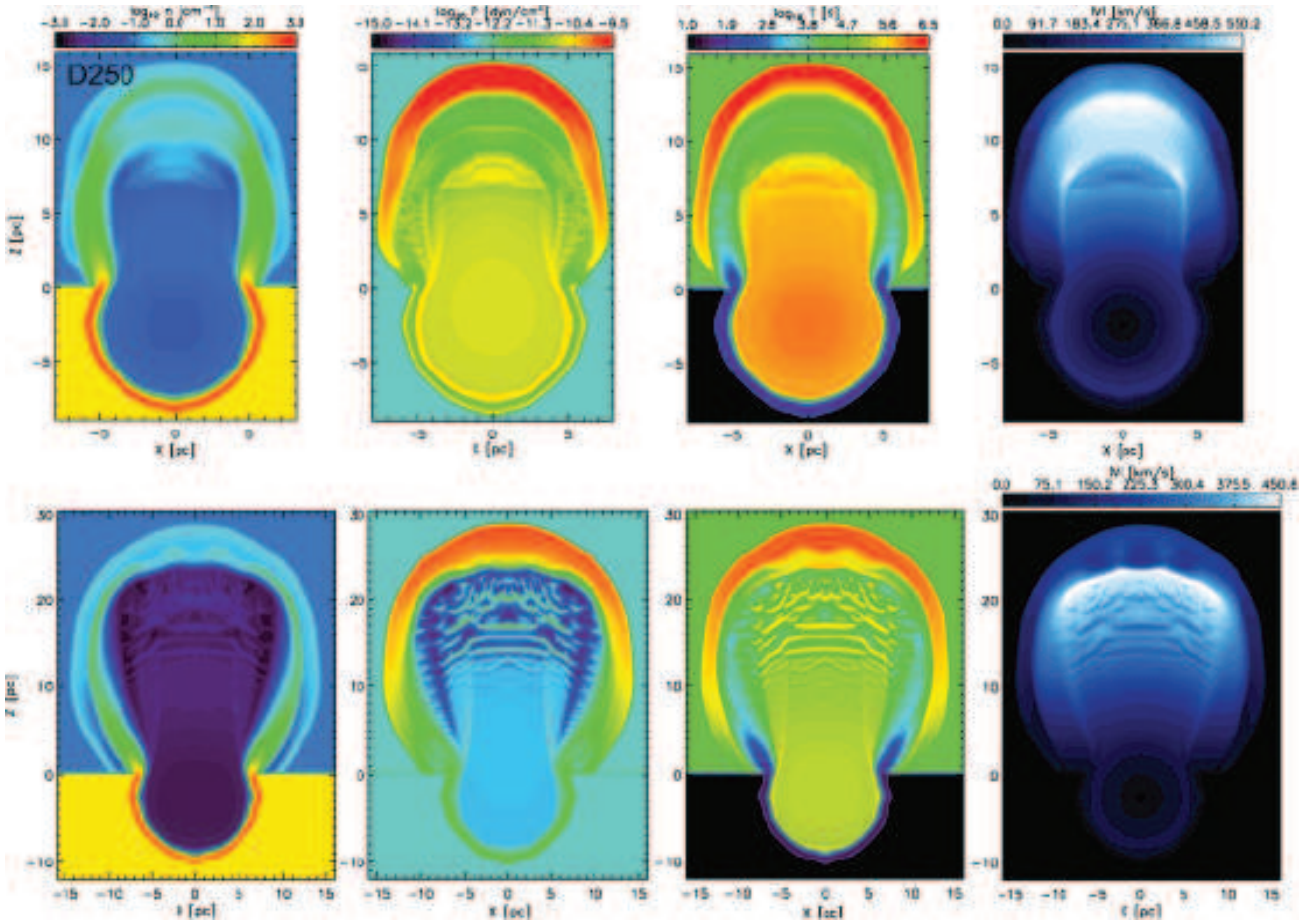}
  \caption{The evolution of model D250. The snapshots of density, pressure, temperature, and speed from left to right columns are drawn at two specific times, i.e., 2.5$\times$10$^{4}$ and 6$\times$10$^{4}$ years after explosion for the upper and lower frames, respectively. The colorbars of density, pressure, and temperature in log scale are shared at each column. To see the overall feature of the remnant, a slice of the quadrant column is copied to the other side with reflective boundary condition assuming symmetry.}
  \end{figure*}

\subsection{Model Parameters}

  We adopt Cartesian coordinates for our 3-D SNR models. Figure~1 shows the schematic description of our models, an SN (red sphere) explodes at a depth, $d$, below the cloud surface between an MC (filled with green color) and the ICM (orange) with a density ratio, $\alpha$, whose remnant will break out through the MC surface and be ejected into the ICM. Each region of the MC and the ICM has uniform density distribution. We vary the density ratio from 10 to 10$^4$ with a fixed hydrogen number density of 100 cm$^{-3}$ and a fixed temperature of 10 K for the MC. In thermal equilibrium, the resulting density of the ICM ranges from 0.01 cm$^{-3}$ to 10 cm$^{-3}$ and the temperature from 10$^5$ K to 100 K. We set the $z-$axis perpendicular to the cloud surface bearing the $x-$ and the $y-$axes. To save computational cost and time, we find solution in one quadrant column along the $z-$axis. Reflective boundary conditions are adopted in the $xz-$ and the $yz-$planes, while continuous boundary conditions in the rest of planes.

  We simulate an SNR from the beginning of the Sedov phase. Because we set the ejecta mass of 10 {\msol} assuming a core-collapse SNR with a massive progenitor, the total mass of the ejecta and the swept-up MC matter becomes 20 {\msol}. The remnant matter is distributed uniformly within a sphere in the radius of 0.89 pc with density of 200 cm$^{-3}$. The total energy of SN explosion is assumed to be 10$^{51}$ erg (E$_{51}$) given in thermal. But in few computational time steps, the physical quantities converge to the Sedov-Taylor solution where the kinetic energy is 28\% \citep{sed46,tay50}. In the case of an SN explosion in a uniform medium, a sharp increase of density appears just behind the shock front in the Sedov-Taylor solution; we term it an \emph{adiabatic} \emph{shell} hereafter. In the snowplow phase of the SNR, the dense neutral radiative shell is formed by cooling in the outermost region of the SNR. The semi-analytic solution of \citet{cio88} modified by \citet{koo04} describes the snowplow phase and determined the radiative shell formation radius, $R_{\rm sf}$, to be 2.78 pc and the shell formation time, $t_{\rm sf}$, to be 2.6$\times$10$^{3}$ years
in a medium with $n_{\rm{MC}}$ of 100 cm$^{-3}$. Thus we can divide our models into two groups according to the SN explosion depths: one case being that an SNR already has a radiative shell in an MC before breakout ($d > 2.78$ pc) and the other case that an SNR breaks out of the cloud surface during its Sedov stage ($d < 2.78$ pc).

Models labelled as Dxxx vary in the depths at which the explosion occurs, while those labelled as Rxxx have varying density ratios. The numeral xxx
trailing the character denotes the depth or the density ratio as indicated in Table 1.
The H032 model uses the highest resolution of 32 [grid/pc] and all models use a computational box of size 16$^{2}$ $\times$ 48 [pc$^{3}$] with a larger height in the ICM to track the evolved stages of the breakout SNR.

Figure 2 shows the time variation of energy (normalized by $E_{51}$), upto an age of 2$\times$10$^{4}$ years in the D250 model. Initially, the SN explosion energy is deposited in thermal energy ($E_{the}$/$E_{51}$ = 1.0) and the remnant follows the Sedov-Taylor solution ($E_{the}$/$E_{51}$ $\sim$ 0.7) in a few computational time step, following the red line which denotes the thermal energy variation. Soon the remnant becomes radiative so that the red line drops sharply. At the same time, the purple line shows the energy loss by the radiative cooling starts to rise more rapidly near the shell formation time, $t_{sf}$. The breakout of the remnant from the MC surface causes the thermal energy to decrease more rapidly due to adiabatic expansion of the escaped part of the SNR into the ICM and the kinetic energy (blue line) to decrease very slowly ($E_{kin}$/$E_{51}$ $\sim$ 0.2-0.3) even after $t_{sf}$. The following sections provide more detail on these calculations.

\section{Results}

\subsection{Standard Model: D250}

  We set the D250 model as a standard representing SNRs breaking through the surfaces of clouds. Especially, the remnant is produced 2.5 pc below the surface of the MC so that it breaks out of the MC in its Sedov stage. We focus on the separation of the adiabatic shell during breakout in the ICM. Figure~3 shows density, pressure, temperature and speed slices of the remnant at two time epochs of 2.5$\times$10$^{4}$ (upper panels) and 6$\times$10$^{4}$ years (lower panels). In Figure~4, the MC and the ICM matter are traced by the coloring method. We label the key structures of the multi-layered structure as \emph{1st} \emph{layer}, \emph{2nd} \emph{layer}, and \emph{ripples}. The \emph{swept-up} \emph{ICM} and the \emph{R-T} \emph{finger} are also labeled.

 \begin{figure}[t]
  \centering
  \includegraphics[width=83mm]{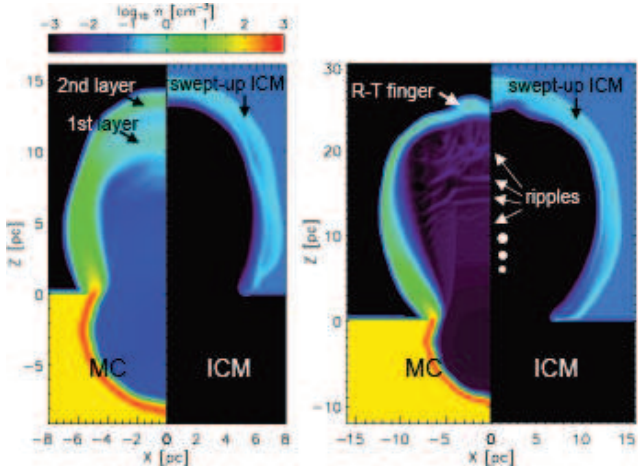}
  \caption{Prominent features of density structures of the model D250. The left side of each frame denotes the MC matter distribution and the right side the ICM matter. The time epochs are 2.5$\times$10$^{4}$ and 6$\times$10$^{4}$ years for the left and right frames, respectively, and the colorbar is the same to that on the density frames in Figure~3.}
  \end{figure}

\begin{figure}[t!]
\includegraphics[width=83mm]{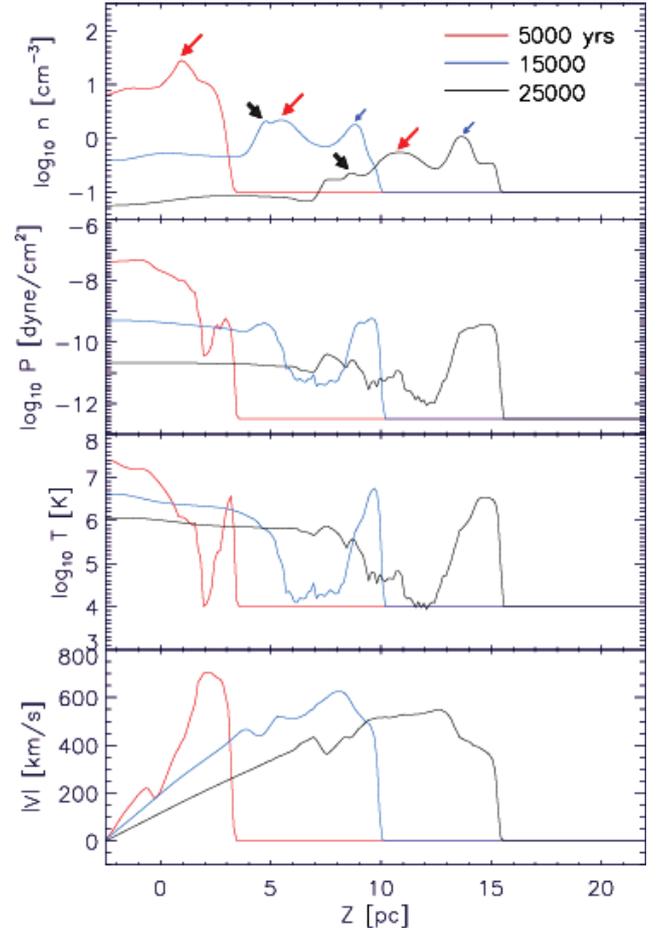}
\caption{One dimensional profiles along the symmetry axis of density, pressure, temperature and speed from top to bottom. Density, pressure, and temperature are drawn in log scale and speed of the matter is drawn in linear scale as a position function of Z, the distance from the MC surface in the unit of parsec. Each frame has three lines of different colors which means the three time epochs of 5$\times$10$^{3}$ (red), 1.5$\times$10$^{4}$ (blue) and 2.5$\times$10$^{4}$ years (black). Arrows in the density frame indicate the first layer (red arrows) and the separated structures (blue and black arrows) from it.}
\end{figure}

Figure 3 shows the general morphology of break-out SNRs such as the spherical radiative shell in the MC and blown-out morphology in the ICM. Even after break-out, the remnant in the MC evolves like an SNR in uniform medium. We see that the density is highest and the pressure and temperature are lowest at the radiative shell due to radiative cooling. But since the cooling effect is small due to the high temperature at the SN explosion site, the temperature is still high at 2.5$\times$10$^{4}$ years.
  But in the ICM, the blown-out part of the remnant makes much more complex structures.
  First of all, we see two green layers of enhanced density in the ICM at 10 and 14 pc, far beyond the MC boundary, which we term 1st and 2nd layer, respectively (see Figure~4).
  Note that the swept-up matter of the first layer is located quite inside the remnant compared with the Sedov-Taylor solution where most of the shocked matter exists just behind shock front. The ripples appear just below the 1st layer.
  The pressure is higher in the 2nd layer and the swept-up ICM matter, while it is much lower around the 1st layer.
  The temperature at the swept-up ICM is maintained higher while the temperature near the explosion site of the SNR goes down rapidly due to radiative cooling.
  Matter speed of the matter is highest at the lower ends of each layer.
  The swept-up ICM matter is colored in bright blue in the density panel, which denotes the shock position propagating into the ICM.

   Before describing the dynamical evolution of the late stages of the remnant, we investigate the process of the separation of the adiabatic shell. We can trace the separated layers as peak positions in 1-D density profile along the symmetry axis. From Figure~5, we note that the separation of the shell occurs due to adiabatic expansion in the following steps. First, the blast wave is accelerated and starts to run away from the adiabatic shell as it breaks out of the MC surface. Second, the pressure between the blast wave and the lagged adiabatic shell has decreased due to the adiabatic expansion (the drop zone shown at 2 pc at 5$\times$10$^{3}$ years in Figure~5 and has been expanding with time). Third, along the pressure gradient, matter at the upper side of the adiabatic shell moves to the Contact Discontinuity between the MC and the ICM (hereafter, CD-MI), piles up, and makes the second layer, pointed out with a blue arrow in the top panel of Figure~5 at 1.5$\times$10$^{4}$ years. Moreover, we can see the beginning of the ripple structure marked with purple arrows in Figure~5.
   The ripples are formed by matter separating from the adiabatic shell due to the same reasons as the second layer: the MC matter from the lower sides of the first layer moves downward and piles up at the contact-discontinuity between the ejecta of the SN and the MC matter. Such chain reactions cause the piled up matter to form small density peaks around the purple arrow in Figure~5 at 2.5$\times$10$^{4}$ years. The downward motion of the ripples create the speed inversions at 4 pc at 1.5$\times$10$^{4}$ years and also at 7 pc at 2.5$\times$10$^{4}$ years.

   Lower frames of Figure 3 show the dynamical evolution at a later, evolved stage of 6$\times$10$^{4}$ years with a single merged shell, Rayleigh-Taylor fingers and ripples. Before 6$\times$10$^{4}$ years, the upper two layers merge into a single shell since the first layer maintains its own speed while the second layer decelerates with outer blast wave. The deceleration results in the R-T instability on the merged layer, shown at 27 pc height in the density frame.
   Since, after merging, the merged shell soon enters the snowplow phase and is decelerated more, the Rayleigh-Taylor unstable structures in the shell grow continuously. They will finally stretch in an upward direction and deform the outermost layer of the swept-up ICM, as discussed in detail in the next subsection.

  \begin{figure}[t!]
  \includegraphics[width=83mm]{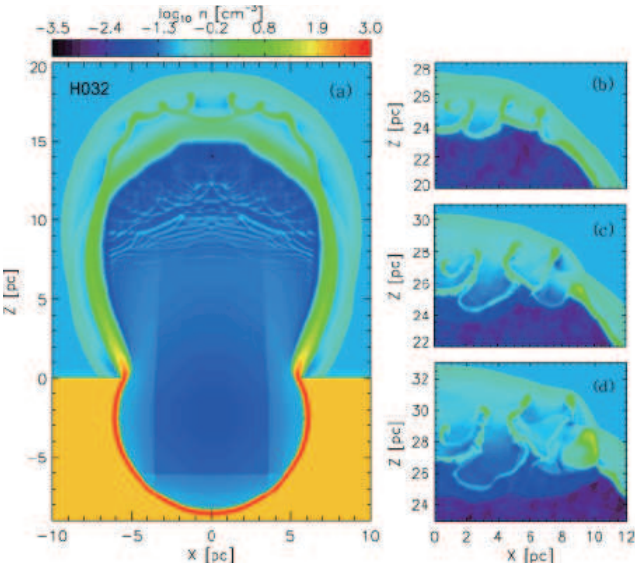}
  \caption{The growth of Rayleigh-Taylor fingers. The left frame labeled (a) presents the overview of density distribution of the remnant of H032 model in late stages and the right frames from (b) to (d) focus on the growth of R-T fingers at the top of the remnant. (a): The first and the second layers are merging at the time of 3$\times$10$^{4}$ years. (b)$\sim$(d): R-T fingers deform the layer of swept-up ICM at 6$\times$10$^{4}$, 7$\times$10$^{4}$, and 8$\times$10$^{4}$ years, respectively.}
  \end{figure}

\subsection{High Resolution Model: H032}

  We need the higher resolution model, H032, to describe the details of the growth of R-T fingers in the late stages of a breakout SNR. The initial conditions of H032 model are the same as those of D250 model but at higher resolution, 32 [grid/pc]. Figure~6 shows the evolution of the R-T fingers and their effects on the environment. The (a) frame captures the moment when the upper two layers are merging.
  We see that the second layer exhibits R-T instability as it approaches the thicker upper layer. After merger (frame b), the fingers are seen to be pushing the CD-MI to deform the outermost layer of the swept up ICM (frame c). Finally, the fingers stretch outward and fragment into several blobs as shown in frame (d) of Figure~5.
  Tenorio-Tagle et. al. (1985), have already argued that fragments arising from the RT instability may be expected to be seen in the ICM if the breakout occurs in the snowplow phase of the SNR. Here, we see that similar fingers grow and become unstable even if the remnant breaks out of the cloud in its Sedov phase.

 \begin{figure*}[t!]
  \centering
  \includegraphics[width=170mm]{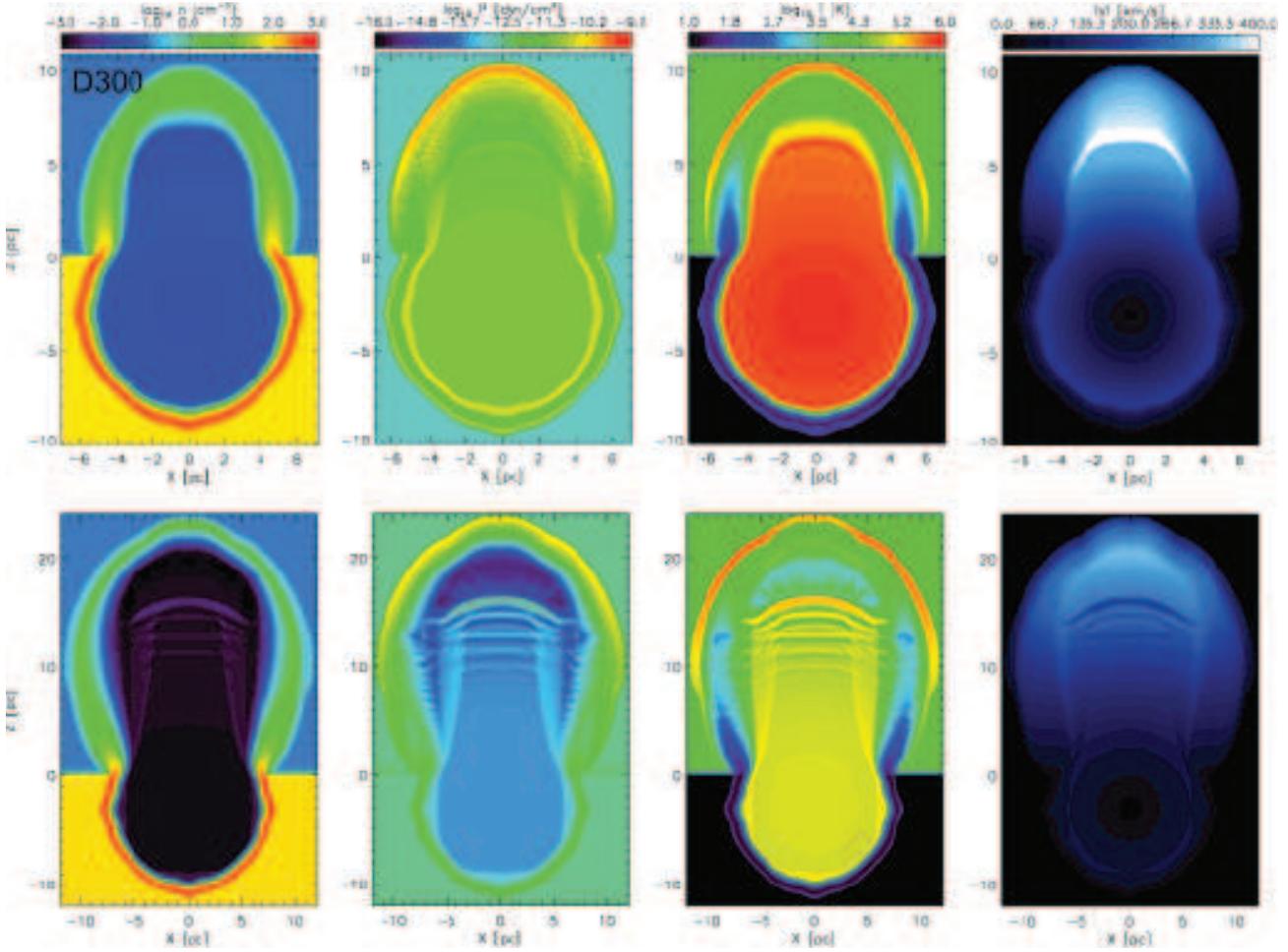}
  \caption{The evolution of model D300. The snapshots of density, pressure, temperature, and speed are drawn from left to right columns. The time epochs are 3$\times$10$^{4}$ (upper) and 8$\times$10$^{4}$ years (lower).}
  \end{figure*}

\subsection{Models with Different Depths:\\ D200, D300, and D350}

 We simulate a group of models with a range of depths, to investigate the influence of the environment of explosion sites. The SNR of D200 model is produced at 2 pc depth below the MC surface, which locate at shallower depth than the radiative shell formation radius, R$_{sf}$ of 2.78 pc, where $n_{\rm{MC}}$ is 100 cm$^{-3}$. Compared with R$_{sf}$, SNRs of D300 and D350 models are produced deeper at 3 and 3.5 pc depths, respectively.

  Figure 7 shows the evolution of the D300 model which does not exhibit the multi-layered structure discussed in the previous section. We can see that the wider radiative shell is elongated in the ICM compared to that inside the MC. The pressure is more uniform compared to the D250 model except in the swept-up ICM layer. And like the D250 model the speed of the matter is fastest at the lower side of CD-EM.
  Because the SNR has already entered the snowplow phase before break-out and its shock speed decreases to a few tens of km/s in the MC, the shock is not accelerated enough to run away from the radiative shell after break-out. Hence the second layer does not appear and the shell retains its shape as a single shell, which is the common characteristic of the models with deep explosion sites. In the lower frames, we can see a deformed shell at the top of the remnant under the effect of the R-T instability similar to the standard D250 model and ripples to move downwards from the shell.

    Figure 8 shows that the SN explosion depth influences the formation of ML structure as well as the overall shape of the remnant in the ICM. In the ICM, the remnant of D350 model shows a single shell and the outer part of the remnant appears more wedge like shape than D200. For D200 model, the number of the upper layers separated from the adiabatic shell is more than that of D250. And we can see the layer just detached from the adiabatic shell has already undergone the R-T instability and shows vertical structures around 10 pc near the symmetry axis. With time, the separate layers merge into a single shell or fade out with ripples. We can see the shape of the outer part of the remnant of D200 is more spherical compared to the other models.

\subsection{Models with Varying Density Ratios:\\ R101, R102, and R104}

These models are simulated with a fixed explosion depth, 2.5 pc and a fixed number density of the MC, $n_{\rm{MC}}$ = 100 cm$^{-3}$. Only the density ratio $\alpha$ between the MC and the ICM is varied with 10$^1$ (R101), 10$^2$ (R102) and 10$^4$ (R104) compared to 10$^3$ (D250).

  Figure 9 shows that the model with higher density ratio shows greater distance between the contact-discontinuities where the multi-layered structure develops. In the upper frames for R101 model, we can see a single shell in the ICM, which is the characteristic density distribution in the ICM of D300 model. Because of the dense ICM, the outer shock cannot run away from the adiabatic shell after breakout. So the adiabatic shell keeps its own shape between the CDs with the multi-layers overlapped. When the shock is decelerated, the shell at the top of the remnant becomes R-T unstable shown in the right frame.
  The multi-layers of the R102 model are very close to each other between 4 and 7 pc heights shown in separated form.
  But they immediately merge into a single shell. Thus, for small alpha, the ML structure is not seen or survives for a short time. Models with larger alpha (D250 and R104) show the ML structure for a longer time compared with the R102 model since the blast wave proceeds faster into the less dense ICM.

  \begin{figure}[t!]
  \centering
  \includegraphics[width=83mm]{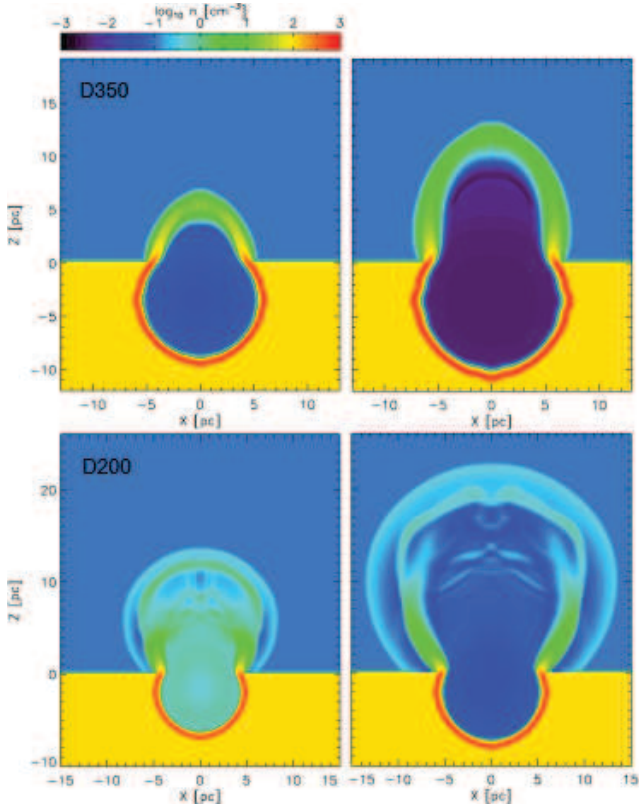}
  \caption{Density snapshots of models D350 (upper) and D200 (lower). The snapshots have different time epochs: D350 at 3$\times$10$^{4}$ and 6$\times$10$^{4}$ years, D200 at 1.5$\times$10$^{4}$ and 3$\times$10$^{4}$ years. All the frames share the same colorbar in log scale.}
  \end{figure}

  \begin{figure}[t!]
  \centering
  \includegraphics[width=83mm]{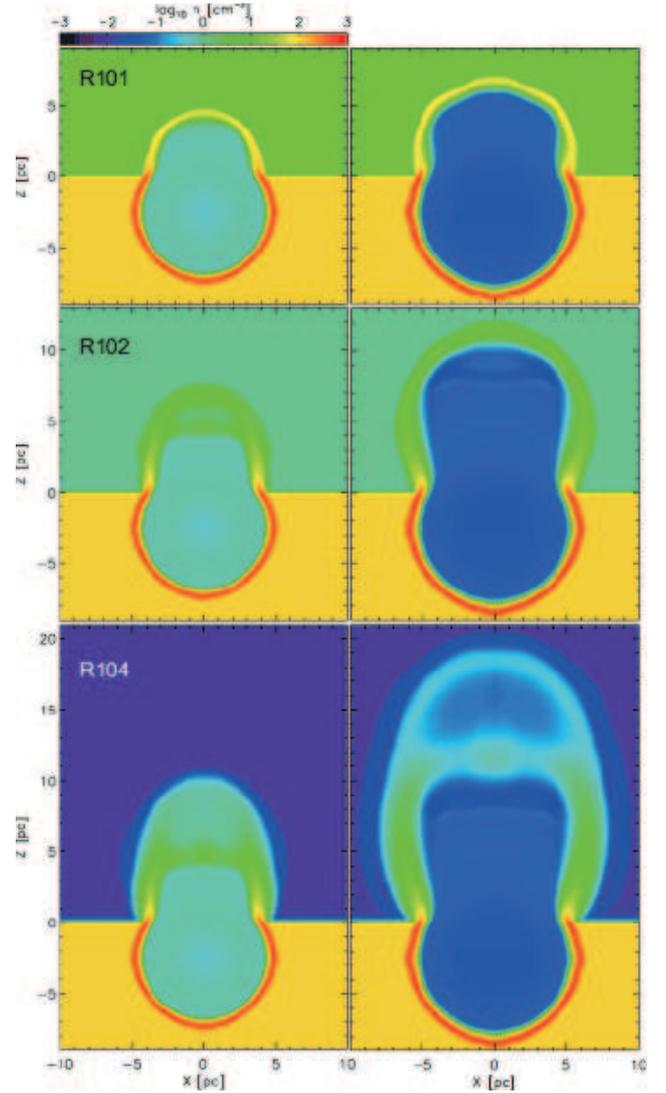}
  \caption{Density snapshots of models R101 (upper), R102 (middle) and R104 (lower) with fixed explosion depths of 2.5 pc. The left frames describes the models of 1.5$\times$10$^{4}$ years and the right frames of 3$\times$10$^{4}$ years.}
  \end{figure}

\section{Discussion}

\subsection{One-Dimensional Experiments}

In this section, we will focus on the propagation of the shock from the breakout SNR in the ICM. To this end, we simplify our 3-D models to 1-D models using spherical coordinates. Because the geometry of the MC surface hardly plays a role in the propagation of the shock along the $z-$axis, we expect that the 1-D models will capture the essential features.
  Since the 1-D model represents the propagation in spherical coordinates, geometric source terms have to be included in the HLL code which is meant for Cartesian coordinates. The 1-D models are solved at higher resolution, 32 [grid/pc] for a time period of 10$^6$ years (ten times longer than 3-D models) at cheaper computation cost. In Figure~10, we display the propagation of the travelling waves along the radial direction inside MCs with radii 3 pc (left panel) and 2.5 pc (right panel) respectively. The density distributions are plotted in logarithmic scale and the colorbars are different in the two pictures to highlight the difference in the density structures.

 In the left frame, the arch is seen to come from the reverse wave which detached from the CD-EM into the MC and rises beyond 1 pc from the explosion center at 4$\times$10$^{3}$ years (upper arrow). As the wave travels in the remnant, it displays the characteristics of a shock such as density increase (seen as a sharp boundary of the arch). When the reverse shock catches up with the outer blast wave in the ICM, the outer shock is pushed out creating yet another reverse shock which again travels towards the center of the remnant. This is seen as a plunging density contrast lasting upto 7$\times$10$^5$ years.

 The right frame shows the evolution of an SNR with an MC radius of 2.5 pc which is smaller than R$_{sf}$. Just after breakout, the adiabatic shell expands in the ICM. We can see the adiabatic shell diffused between yellow sharp lines: the upper line for the swept-up ICM and the lower one for the reverse shock. The inclined arch of the reverse shock rises again at 4$\times$10$^{3}$ years, but falls by 3$\times$10$^5$ years, which is much earlier than that in the former case, since the pressure in outermost region is higher.

 \begin{figure*}[t!]
  \centering
  \includegraphics[width=160mm]{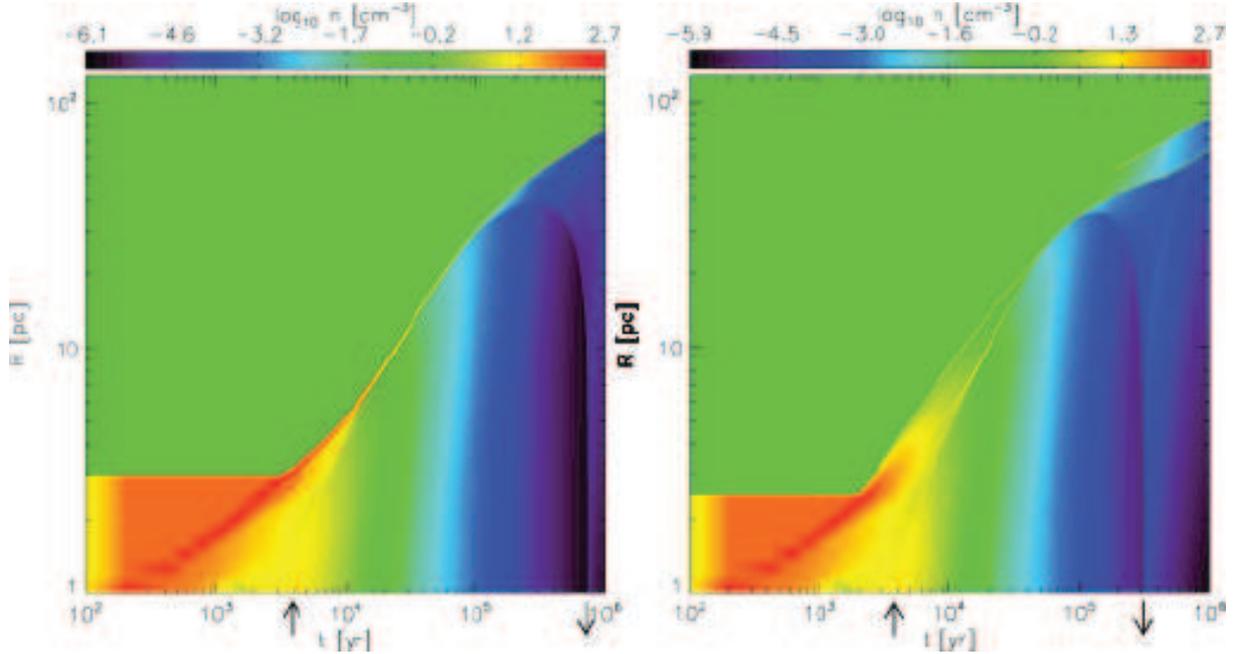}
        \caption{Radial density profiles of 1-D SNR simulations as a function of time. The left frame shows the density profiles where the supernova explodes at the center of an MC sphere with a 3-pc radius, while MC with a 2.5-pc radius in the right frame. The arrows are marked at the bottom of each arch to denote the rise and the fall of the reverse shock inside the remnant.}
  \end{figure*}

  From Figure 11, we might suspect that the shock propagation may be modelled by piecewise straight lines, both in the 1-D (left) and the 3-D case (right).
The shock position is determined by the temperature peak just behind the shock front for 1-D models and along the $z-$axis for 3-D models.
 In the left frame, the shock positions are plotted in log scale for four 1-D simulations with different radii from 2 to 3.5 pc. The upper two lines with smaller MCs (2.0 and 2.5 pc radii, smaller than R$_{sf}$) show similar slopes. Inside the MCs, they follow the Sedov-Taylor solution with a slope of $2/5$ \citep{sed46}. The slopes increase suddenly to $3/4$ after breakout and decrease to $3/10$ after the radiative cooling becomes dominant. For the lower two lines, the slopes of remnants with larger MCs (3.0 and 3.5 pc radii, larger than R$_{sf}$) are changed by the arrival of the reverse shock. Inside MCs, the remnants follow the Sedov solution and soon enter the snowplow phase. The slope of the blast wave increases to $3/4$ with breakout and, once the reverse shock catches up with the outer shock, the slope increases to $4/5$. The remnants in the ICM finally enter the snowplow phase again and the slopes follow $3/10$. In the right frame, the shock propagations of 3-D models show similar trends to those of 1-D models, but we can check the shock propagations only at early stages due to the limit of short computational period. Figure~12 shows the schematic descriptions for the shock evolution, and is meant to be a simplified representation of Figure~11. We can see four kinds of slopes at each plot from the Sedov phase ($2/5$), the snowplow phase ($3/10$), the breakout ($3/4$), and the reverse shock ($4/5$).

  \begin{figure}[t!]
\centering
\includegraphics[width=83mm]{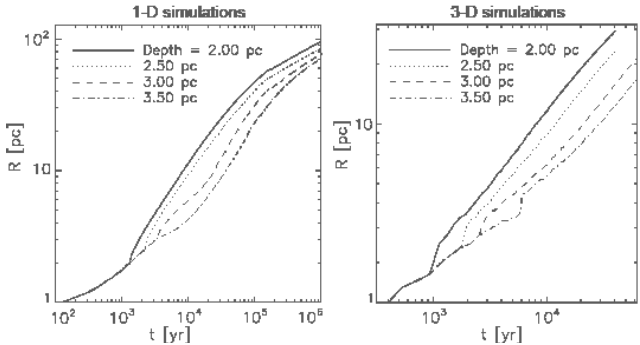}
\caption{Shock propagations of 1-D and 3-D simulations for $\alpha$ = 1$\times$10$^{3}$ with different radius of MC: 2.0 pc (solid line), 2.5 pc (dotted line), 3.0 pc (dashed line) and 3.5 pc (dot-dashed line).}
\end{figure}

\begin{figure}[t!]
\centering
\includegraphics[width=83mm]{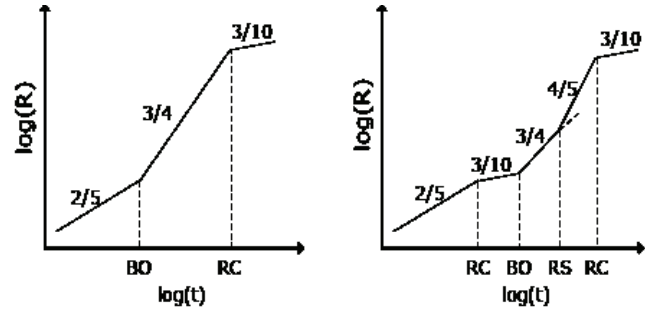}
\caption{Schematic plots of shock position with time with different radii of MCs in logarithmic scale from the shock propagations of 1-D cases with the left frame of Figure~11. The broken lines describe the shock propagation of the models that an SN explodes at a shallow depth (left) and that an SN explodes deep inside an MC (right). The numbers over lines denote the powers. The vertical dashed lines mark the points of specific events such as breakout (BO), dominance of radiative cooling (RC) and arrival of the reverse shock (RS).}
\end{figure}

In Figure 13, we compare the density distribution along the $z-$axis of the H032 model with the previous 1-D models and also compare with analytic solutions that describe the reverse shock. With the MC radius set to 2.5 pc, the arch of the reverse shock appears at 4$\times$10$^{3}$ years (this is the same as in the right panel of Figure~10 indicated by the upward arrow).
  The dotted line in Figure~13 shows the position of the shock obtained from the 1-D simulation and are seen to be very close to those obtained from the H032 model (the boundary of the blue region). Also shown is the semi-analytic solutions obtained by \citet{koo90} for the shell (dashed line) and shock (dot-dashed line). The semi-analytic solutions differ because they describe the evolution in an exponential medium unlike in the current work which has a jump in the density. However, the position of the shell is seen to reasonably capture the evolution of the first layer. This is because most of the swept-up matter is left in the first layer after breakout.

  \begin{figure}[t!]
  \centering
  \includegraphics[width=83mm]{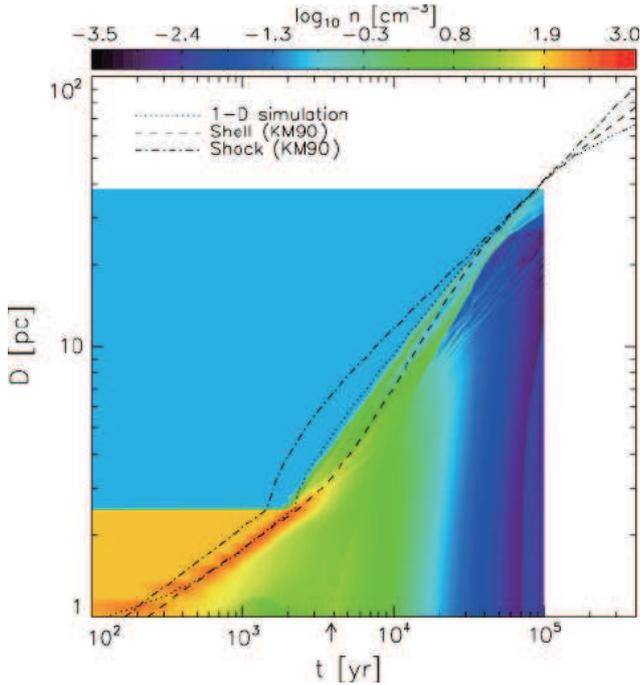}
  \caption{One-dimensional density profiles of H032 along the $z-$axis with time (background) and shock positions in 1-D simulations (dotted line). The analytic solutions of  \citet{koo90} for the shell (dashed line) and the shock (dot-dashed line) are plotted on the profiles. We put the arrow under the time axis to mark the start of arch, which denotes the reverse shock travelling inside the remnant.}
  \end{figure}

   \begin{figure*}[t!]
  \centering
  \includegraphics[width=160mm]{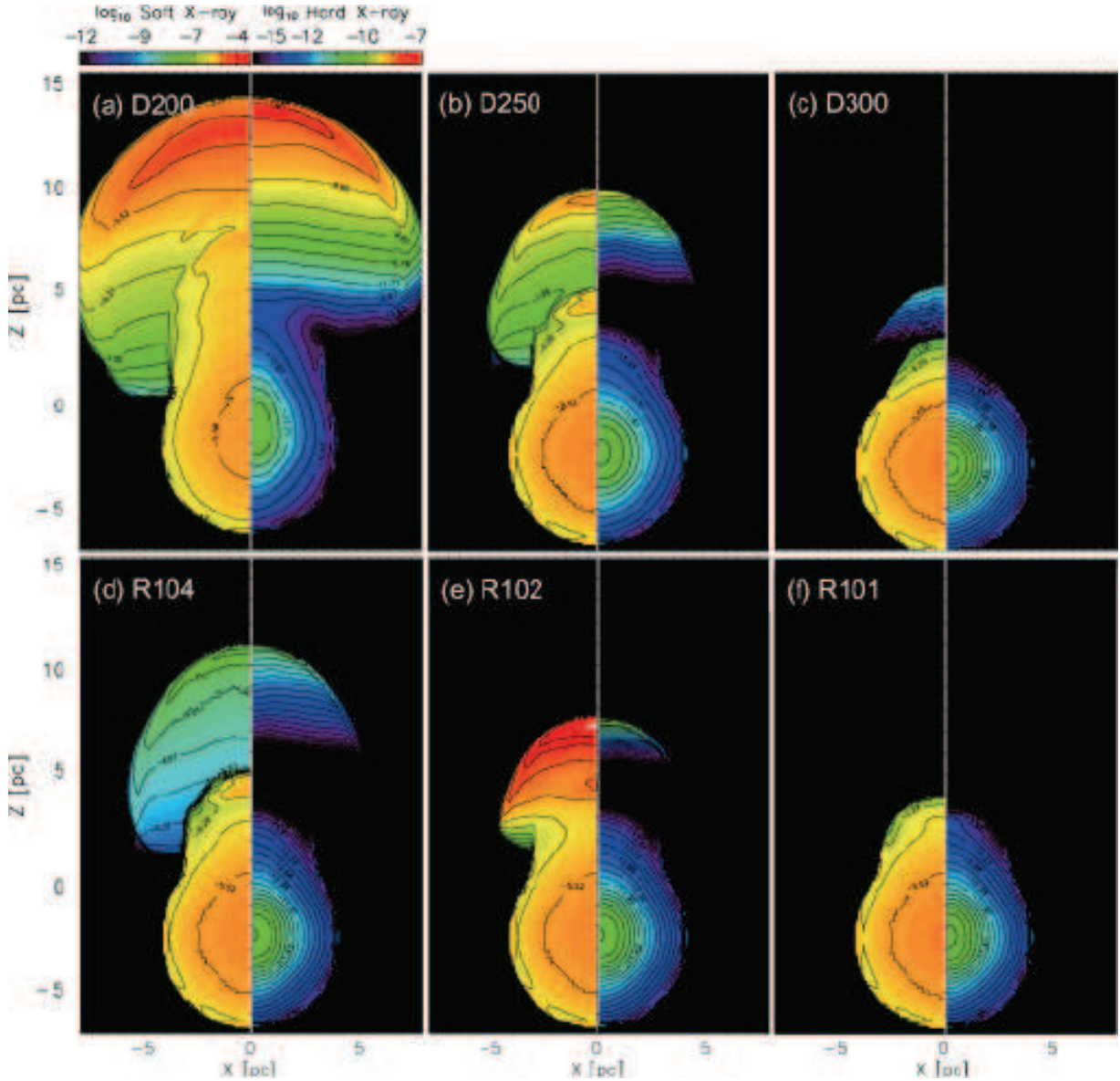}
    \caption{Simulated X-ray surface brightnesses in soft (left) and hard (right) X-rays of the models. The colorbar over the left part of frame (a) represents the soft X-ray surface brightness and that over the right part shows the hard one in logarithmic unit of erg/cm$^2$/s/sr. The time epoch for the frames is 1.5$\times$10$^{4}$ years.}
  \end{figure*}

  \begin{figure*}[t!]
  \centering
  \includegraphics[width=170mm]{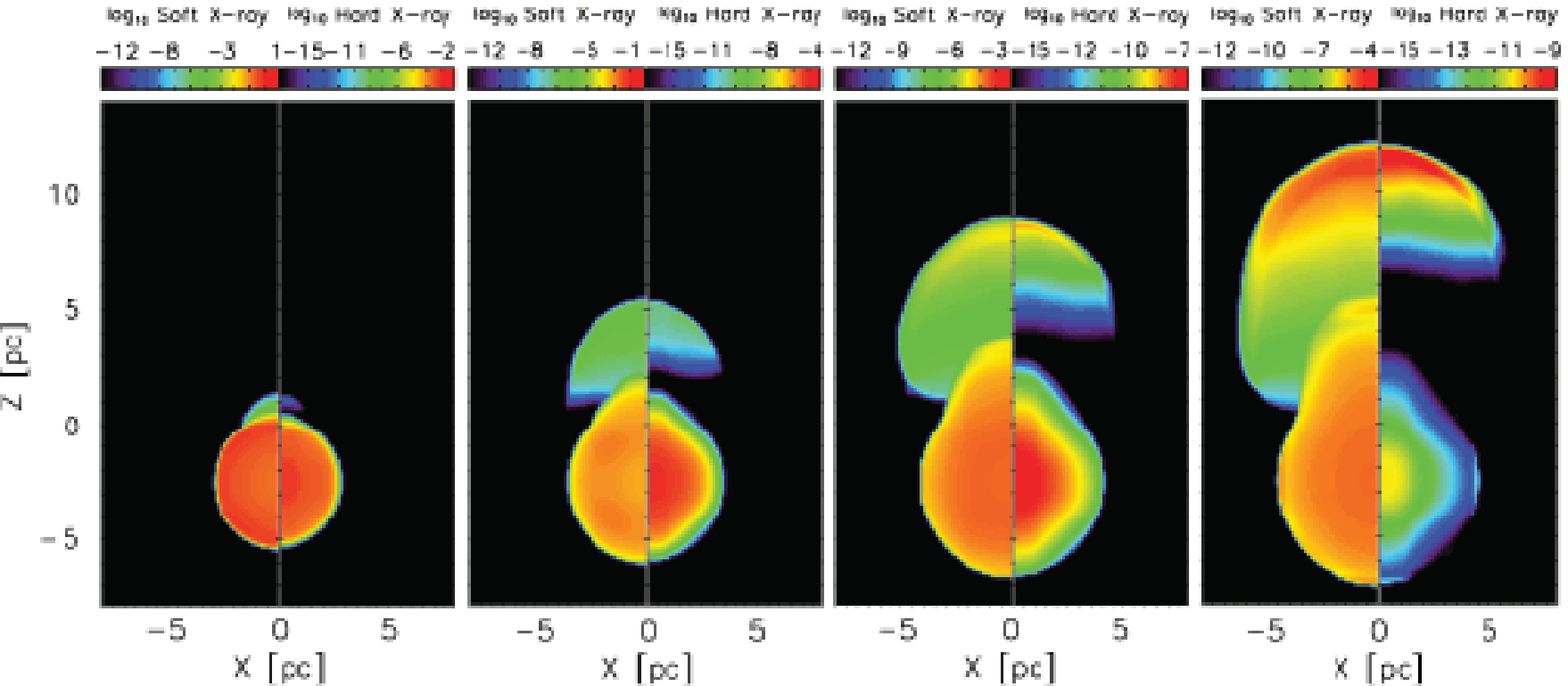}
    \caption{Snapshots of the X-ray surface brightness of model D250. Each frame shows the brightness in soft (left part) and hard (right part) X-ray bands in logarithmic units of erg/cm$^2$/s/sr. The colorbars are set to show the full dynamical range of the brightness at each time epoch of t = 2.5$\times$10$^{3}$, 7.5$\times$10$^{3}$, 1.25$\times$10$^{4}$, and 1.75$\times$10$^{4}$ years (from left to right frames). The bright part at the rim of the remnant inside MC at 1.75$\times$10$^{4}$ years would be artifact since we cannot resolve the contact discontinuity sharply by HLL code.}
  \end{figure*}

\subsection{X-ray Morphology}

In this section, we will simulate the surface brightness in X-ray of the MM-SNR to study their morphology since these are diagnosed by showing center-bright
thermal X-ray emission while being shell-bright in radio. The X-ray brightness of our model is calculated using the emissivity tables of \citet{ray77}, which includes recombination, bremsstrahlung, two-photon processes and line emissions. The chemical abundance is taken to be the solar abundance of \citet{and89}.
The X-ray emissivity is multiplied by $n_{\rm{e}}n_{\rm{H}}$ ($n_{\rm{e}} = 1.2 n_{\rm{H}}$) and divided by 4$\pi$ to get the emission per unit solid angle. Using the optically thin approximation, the simulated brightness on the $xz-$plane is obtained by integration along the line of sight.

Figure 14 shows the simulated X-ray brightness of our models in the soft (0.1-3.0 keV, left panel) and the hard (3.0-8.0 keV, right panel) X-ray bands at 1.5$\times$10$^{4}$ years. The upper panels are for varying depths of the explosion, while the lower panels capture the variation with the density ratio. In all cases, the X-ray morphology of the SNRs in the radiative stage resembles a mushroom with a cap in the ICM and a stem through an MC.

Based on Figure 14, we can say that the center-bright X-ray morphology can arise from an SNR which explodes inside a dense medium without invoking any thermal conduction mechanism \citep{til06}. During the early stages of SNR evolution, the soft X-ray brightness tracks the shell-bright morphology, since newly swept-up MC matter ensures that the density and temperature are highest near the shock position (see also Figure~15). However at late stages, the material in shell region cools below 10$^4$K which is too low for X-ray emission, but that in explosion site could still be hot enough to continue emitting bright X-rays inside the MC. We can check the center-bright morphology in soft X-ray band of SNRs inside MCs in Figure~14 and the third and the last frames in Figure~15, while the hard X-ray brightness shows the center-bright morphology from the earlier to the late stages.

In the ICM, otherwise, we can see the shell-bright morphology in SNRs in X-rays in Figure 14. Since the blast wave is re-accelerated by breakout from the MC surface to stay in adiabatic state, the SNR has maximum density and temperature around shock toward the ICM. Especially, the hard X-ray can point out the shocked ICM described a thin layer above CD-MI, while the soft X-ray brightness is shown more diffuse. In the (c) and (f) frames, the cap parts are fading out since the cooling effect is dominated.

  The ML structure can also contribute to central-brightening in soft X-rays, as remarked earlier. We notice that soft X-ray emission traces the multi-layered structure in Figure~14. Since the MLs can retain high density inside an SNR after it breaks out through the MC, they can supply enough matter to enhance the surface brightness. In (a), (b), (d), and (e) frames, we can see the bright soft X-ray on the multi-layers. The (a) frame shows several bright X-ray features from 5 pc to 13 pc height along the symmetry axis. In the other frames, only the first and the second layers emit bright soft X-rays unlike (a) frame, because the temperature has been lowered rapidly by adiabatic expansion between the two layers. The soft X-ray of the second layer depends on the ICM density. Moreover, under specific conditions such as higher resolution or larger density ratio, these layers could be distorted or broken apart by R-T instability to form clumpy structures inside the remnants.

  There are three prominent points in the simulated X-ray surface brightness of our models. First, the ML structure enhances the soft X-ray brightness, which may reveal the clumpy and complex internal X-ray structures of MM-SNRs. Second, the center-bright morphology in soft X-ray can be formed in the evolved phase of models inside a dense medium. Third, a breakout SNR shows the shell-bright X-ray morphology toward the ICM.

\subsection{3C 391}

3C 391 is a prototype of MM-SNRs \citep{rho98}.
In radio continuum, we can see its blown-out morphology clearly across the surface of an MC.
The remnant is elongated from northwest to southeast and shows a bright rim inside the northwestern MC
indicating the interaction with the remnant \citep{rey93}.
On the other hand, the $Chandra$ images \citep{che01,che04,che05} show
that clumpy X-ray emission of thermal origin is filling the inside of the remnant (see the upper image of Figure~17). We simulate an additional 3-D model to reproduce the X-ray morphology of 3C 391. From the X-ray images of \emph{Einstein} \citep{wan84} and \emph{Chandra} observations of \citet{che04}, we assume that the SN explodes around 2.4 pc depth (1$'$ at 8 kpc distance) under the surface of a dense MC, where we set the hydrogen number density of the MC, 40 cm$^{-3}$, as a upper limit from \citet{che01}. Then the radiative shell formation time and radius become t$_{sf}$ = 4.4$\times$10$^{3}$ years and R$_{sf}$ = 4.1 pc. The age of SNR 3C 391 may be estimated as 8.5$\times$10$^{3}$ years based on the assumed hydrogen density of the MC and the distance from the explosion site to the dense shell of the remnant toward the inside of the cloud, 5.1 pc (2.2$'$), using the semi-analytic solution of \citet{cio88} as modified by \citet{koo04}. For the other initial conditions, we set the density of the ICM of 0.1 cm$^{-3}$ as a typical value of the ICM and the resolution of 8 grids a pc.

  Figure 16 shows the density and temperature distributions (left) and the X-ray surface brightness (right) of a new model at 8.5$\times$10$^{3}$ years from SN explosion.
   Considering the lower density of the MC of a new model compared to that of the standard one,
   we may expect that the part of the remnant inside the MC still keeps its shell-bright morphology in soft X-ray band.
   Furthermore, we notice that the evolved ML structure includes R-T fingers and fragments around 5 $\sim$ 9 pc heights around the symmetry axis in the left frame. These fingers and fragments might be the origin of the clumpy structures inside 3C 391, but they are not reflected in the X-ray brightness due to the projection effect and also they cannot enhance the X-ray emission enough to explain the bright clumps of the regions of 6, 7, and 8 in the top image of Figure~17.

  \begin{figure*}[t!]
  \centering
  \includegraphics[width=160mm]{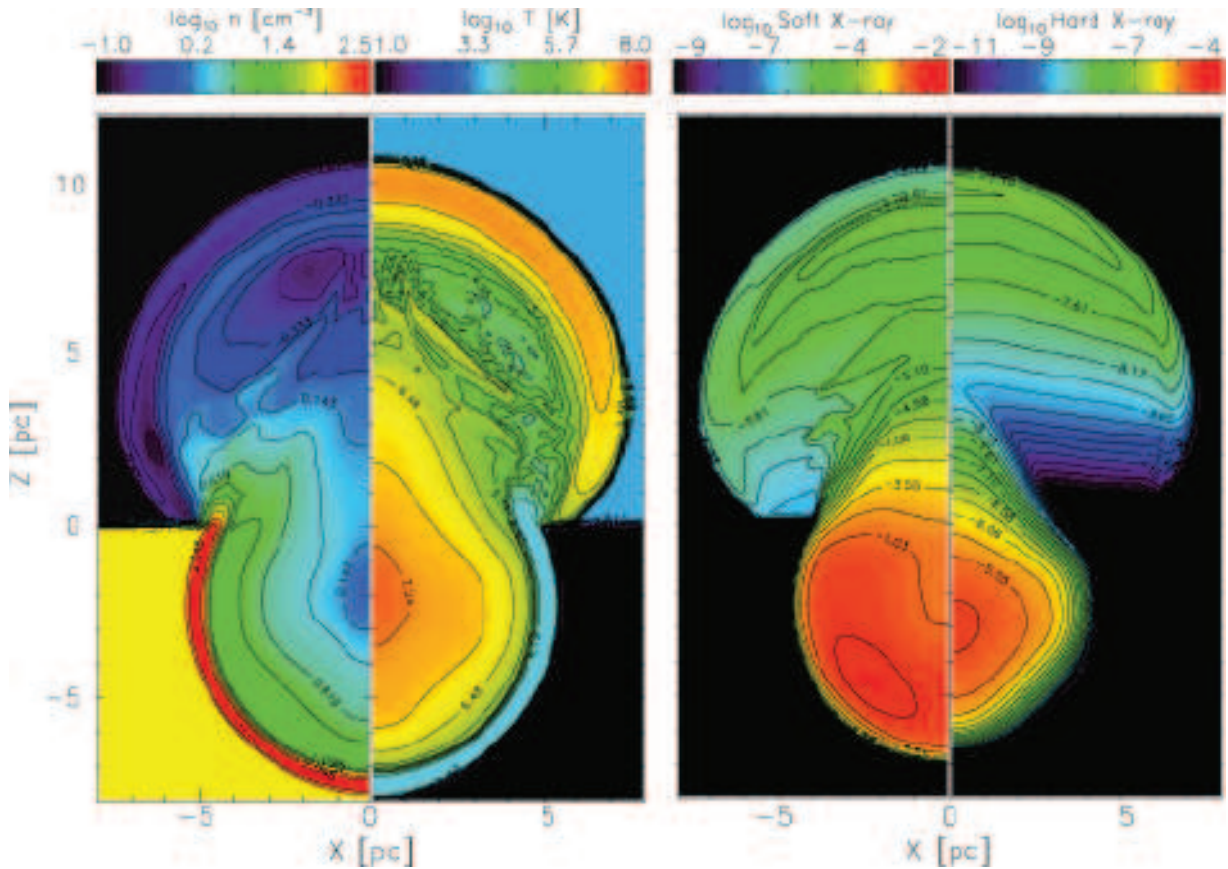}
    \caption{Density and temperature snapshots (left frame) and the soft and hard X-ray surface brightnesses (right frame) of the model for the MM-SNR 3C 391 in the same unit as Figure~14. The time epoch is 8.5$\times$10$^{3}$ years. Each colorbar follows log scale and the unit of the X-ray brightness is erg/cm$^2$/s/sr.}
  \end{figure*}

    \begin{figure}[t!]
    \centering
  \includegraphics[trim=0mm 0mm 25mm 0mm, clip, width=83mm]{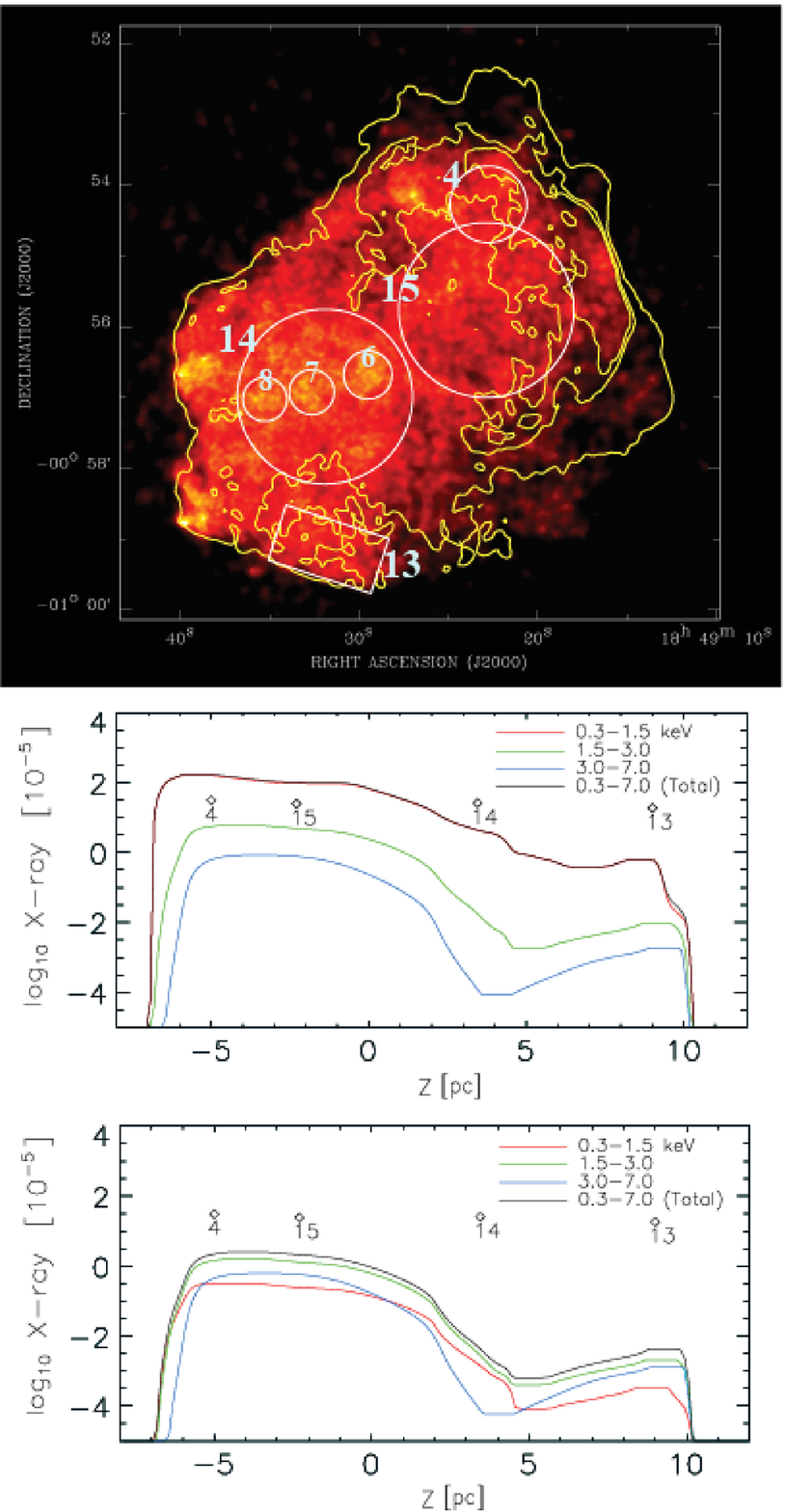}
    \caption{(top image) $Chandra$ X-ray (0.3-7.0 keV) image \citep{che04} with the FIRST 1.4 GHz radio contours (Becker et al. 1995) of the SNR 3C 391. (middle frame) X-ray brightness profiles along the symmetry axis from our model, where red, green, blue, and black curves are for 0.3-1.5, 1.5-3.0, 3.0-7.0, and 0.3-7.0 keV bands, respectively. The diamond symbols represent the observed brightnesses in
3C 391 which are from \citet{che04}. They are labeled by the region numbers in \citet{che04}. The unit of the X-ray brightness is erg/cm$^2$/s/sr. The red and black lines are almost overlapped, since most of X-rays are radiated in the 0.3-1.5 keV.
(bottom frame) Same as the middle frame but including the attenuation by foreground and MC media (see text for details). }
  \end{figure}

We show the image of the $Chandra$ X-ray observation with contours of the VLA FIRST (Very Large Array Faint Images of the Radio Sky at Twenty-cm) 1.4 GHz radio continuum \citep{bec95} of 3C 391 and profiles of the calculated X-ray brightness along the axis of symmetry in Figure~17. In order to compare the observed brightness of the remnant to our simulated X-ray brightness, we select four different regions of 4, 13, 15, and 14 from \citet{che04} in the top image of Figure~17, which represent the regions near the shock front toward the MC and the ICM and the inner regions of the remnant in the MC and the ICM, respectively. We mark the observed X-ray brightness of the regions with diamond symbols on the lower frames of Figure~17. If there is no attenuation of X-rays (middle frame), most of the X-rays are emitted in the soft band (0.3-1.5 keV). However, since the X-ray emitting hot gas is surrounded by the dense MC material, we need to consider the attenuation of X-rays due to the MC and the foreground media (bottom frame). According to \citet{che04}, the column density to the SNR is 2.9$\times$10$^{22}$ cm$^{-2}$ to the ICM area and 3.5$\times$10$^{22}$ cm$^{-2}$ to the MC area. We could expect that extinction will decrease the overall brightness significantly but would not change the morphology considerably since the column density difference between the MC and the ICM areas is only 0.6$\times$10$^{22}$ cm$^{-2}$. We compute the attenuated X-ray emission using the energy-dependent transmission curve of \citet{sew99} with the column densities of \citet{che04}. In detail, we calculate the flux weighted mean transmission with temperature for the column densities in the specific X-ray energy bands. Then we multiply it to the X-ray emission on each grid and integrate the transmitted emission toward the line of sight. The result is shown in the bottom frame of Figure~17.

$Chandra$ X-ray image of 3C 391 shown in the top image of Figure 17
 is somewhat different from our simulated X-ray profiles shown in the lower frames.
  The SNR has an almost uniform X-ray brightness
    while our simulation shows
    that the brightness in the ICM is much fainter than that in the MC region.
  Toward the SNR bubble within the MC, the emission is from relatively dense ($\sim$ 5 cm$^{-3}$) hot gas near the MC,
    while toward the SNR in the ICM,
    it is mostly from the gas in the adiabatic shell where the density is much lower ($\sim$ 0.5 cm$^{-3}$) (see Figure~16).
  The calculated emission is brighter than the observed for the SNR part in the MC area
    while it is fainter for the SNR part in the ICM in the middle frame of Figure~17.
  But if we assume extinction, the total brightness is fainter than the observed even toward the SNR part in the MC
    (see the bottom frame of Figure~17), and now the dominant emission becomes from 1.5-3.0 keV band (green curve).
  As it is easily expected, however, the overall shape of profiles is not changed much.

We suggest that thermal conduction and evaporation of preexisting
cloudlets can explain the difference between the $Chandra$ X-ray observational result
and our results.
  For the SNR part embedded in the MC, the interface between the hot gas and the MC might be subject to conduction, which will increase the gas density in the hot bubble \citep{cox99}. A factor of 3 increase in gas density can explain a factor of 10 difference in brightness since the brightness is roughly proportional to the square density.
  On the other hand, for the SNR part in the ICM, the difference in the X-ray brightness is more than three orders of magnitude in the bottom frame of Figure~17,
  which might be difficult to be explained in a large scale conduction alone.
The $Chandra$ image suggests that the medium is clumpy with dense cloudlets.
The density is, for example, 5-7 cm$^{-3}$ at the regions of 6-8 in the top frame of Figure~17 whereas our characteristic density of those regions is $\sim$0.5 cm$^{-3}$ (Figure~16).
Therefore, there were likely dense clumps in the past, which might have provided additional mass to the hot gas by thermal evaporation \citep{whi91}. The increase of gas density from such evaporation of clumps together with the increase of temperature by thermal conduction from the hotter gas within the bubble might explain the difference. Such scenario may be explored in a future study.

  \section{Summary and Conclusions}

  We have simulated breakout morphology SNRs with different explosion depths and density ratios to show the evolution of SNRs breaking through molecular clouds (MCs). We have presented a fiducial model where the explosion depth is 2.5 pc below the surface of an MC, which breaks out the surface in its Sedov phase. The outermost shell in the Sedov phase, which we call the adiabatic shell, is separated after the breakout in two thick layers at its upper side and ripples at its lower side, which we call the \emph{multi-layer} structure. It is noticeable that the shocked ambient matter can exist inside a remnant in the form of the ML structure, which cannot be expected in Sedov-Taylor solution.

  The environmental effect on the evolution the breakout SNRs is also investigated. When an SNR is produced closer to the MC surface, the number of layers increases at the front of the original adiabatic shell. Also in the more rarefied intercloud medium (ICM), the ML structure survives longer time. If the radiative shell is formed before breakout, we cannot see the ML structure because the outer shock is not fast enough to run away from the radiative shell.

  The growth of R-T fingers is another key point in the paper. If the SNR breaks out of the MC surface at its Sedov stage, the R-T fingers grows on the layers in the front side of the original adiabatic shell at the beginning of the ML structure. The simulation with highest resolution shows the evolution of R-T fingers at the top of the remnant after merging of the multi-layers, which are fragmented into several blobs and penetrating the shock front.

  We have discussed the shock propagation with simplified one-dimensional (1-D) models. We have noticed that the slope of shock position as a function of time is influenced by the reverse shock, detached from the dense shell in the MC, to rise to $4/5$, while the other slopes are denoted as $3/4$ due to breakout, $2/5$ from Sedov-Taylor solution, and $3/10$ due to radiative cooling. The shock propagations of 3-D models are well described by the simplified 1-D models and the adiabatic shell of the ML structure is fitted well with the semi-analytic solutions of shell in \citet{koo90} for a spherically symmetric blast wave on the ambient medium with density drop.

  From the simulated X-ray brightness of our models, three key points can be inferred: (1) The ML structure can enhance the soft X-ray brightness (2) A remnant in an MC can appear centrally-brightened in X-rays around the explosion site in the evolved phase (3) The newly swept-up ICM matter emits hard X-ray at the swept-up ICM behind the outer shock.
  Compared with the $Chandra$ images of 3C 391 of \citet{che04} as a fiducial mixed morphology SNR, the simulated surface brightness is consistent in X-ray brightness and the transmitted X-rays of 3C 391 in the MC quantitatively. But we can see difference in X-ray brightness between the model and $Chandra$ observation, which cannot be fully explained by our simplified models without thermal conduction or preexisting cloudlets outside the MC.

\section{Acknowledgements}
  This research was supported by Basic Science Research Program through the National Research Foundation of Korea(NRF) funded by the Ministry of Science, ICT and future Planning (2014R1A2A2A01002811). Numerical simulations were performed by using a high performance computing cluster in the Korea Astronomy and Space Science Institute. We wish to thank Dr. Chen, Y. for providing $Chandra$ images of 3C 391.


\end{document}